\documentclass[%
 reprint,
superscriptaddress,
showpacs,preprintnumbers,
nofootinbib,
 amsmath,amssymb,
 aps,
 twocolumn,
]{revtex4}

\usepackage{graphicx}
\usepackage{dcolumn}
\usepackage{bm,braket}

\input{colordvi.tex}
\begin{document}

\preprint{-}

\title{Cosmic microwave background bispectrum of vector modes induced from primordial magnetic fields}
\author{Maresuke Shiraishi}
\affiliation{Department of Physics and Astrophysics, Nagoya University,
Aichi 464-8602, Japan}
\email{mare@a.phys.nagoya-u.ac.jp}
\author{Daisuke Nitta}
\affiliation{Department of Physics and Astrophysics, Nagoya University,
Aichi 464-8602, Japan}
\email{nitta@a.phys.nagoya-u.ac.jp}
\author{Shuichiro Yokoyama}
\affiliation{Department of Physics and Astrophysics, Nagoya University,
Aichi 464-8602, Japan}
\author{Kiyotomo Ichiki}
\affiliation{Department of Physics and Astrophysics, Nagoya University,
Aichi 464-8602, Japan}
\author{Keitaro Takahashi}
\affiliation{Department of Physics and Astrophysics, Nagoya University,
Aichi 464-8602, Japan}

\date{\today}

\begin{abstract}

We calculate the cosmic microwave background bispectrum of vector modes induced
from primordial magnetic fields.
We take into account the full angular dependence of the bispectrum
and discuss the amplitude and also the shape of the bispectrum in $\ell$ space. 
In the squeezed limit as $\ell_1 = \ell_2 \gg \ell_3$, we estimate a typical values of the normalized reduced bispectrum as 
$\ell_1(\ell_1 + 1)\ell_3(\ell_3+1)|b_{\ell_1\ell_2\ell_3}| \sim 2 \times 10^{-19}$, for the strength of the primordial magnetic field smoothed on $1 {\rm Mpc}$ scale $B_{1 \rm Mpc} = 4.7 \rm nG$ assuming a nearly scale-invariant spectrum of magnetic fields. We find that 
a constraint on the magnetic field strength will be roughly estimated as
$B_{1 \rm Mpc} < 10 {\rm nG}$ if PLANCK places a limit on the nonlinearity parameter of local-type configuration as $|f^{\rm local}_{\rm NL}| < 5$.

\end{abstract}

\maketitle


\section{Introduction}\label{sec:intro}

Non-Gaussianity of primordial curvature perturbation has been recently
focused on as a powerful tool to probe the mechanism of generating
the seed of structures in our Universe~\cite{Komatsu:2010hc}. Current observations of the cosmic microwave background (CMB) anisotropies
(and also large scale structure) indicate
that the statistics of the primordial
curvature perturbation is almost Gaussian. Although
it is difficult to detect the primordial non-Gaussianity
only by analyzing the power spectrum,
the bispectrum of the primordial fluctuation is an useful observable 
to detect the non-Gaussianity for such weakly
non-Gaussian fluctuations.
For example, 
the nonlinearity parameter $f_{\rm NL}$,
which is defined as~\cite{Komatsu:2001rj}
\begin{eqnarray} 
\Phi = \Phi_{\rm G} + f_{\rm NL} \left( \Phi_{\rm G}^2 - \langle \Phi_{\rm G}^2\rangle \right)~,
\label{eq:localfNL}
\end{eqnarray}
where $\Phi$ is primordial curvature perturbation and $\Phi_{\rm G}$ denotes the Gaussian part,
has been commonly used 
for discussing the size of the non-Gaussianity.
The bispectrum of $\Phi$ is given by
\begin{eqnarray}
\begin{split}
\langle \Phi({\bf k}_1)\Phi({\bf k}_2)\Phi({\bf k}_3)\rangle 
&= (2\pi)^3 F(k_1,k_2,k_3)\\
&\quad \times \delta \left({\bf k}_1+{\bf k}_2+ {\bf k}_3 \right)~, \\
F(k_1,k_2,k_3) &= 2 f_{\rm NL} \left( P(k_1)P(k_2) + {\rm perms.} \right)~. \label{eq:local_pri_bis}
\end{split}
\end{eqnarray}
This means that the nonlinearity parameter $f_{\rm NL}$ characterizes
the size of the bispectrum of the curvature perturbation.
In recent years,
depending on the shape of the bispectrum,
three types of $f_{\rm NL}$ have been mainly discussed: the local type $f_{\rm NL}^{\rm local}$, equilateral type $f_{\rm NL}^{\rm equil}$, and orthogonal type $f_{\rm NL}^{\rm orthog}$\footnote{
Local type nonlinearity parameter 
is corresponding to the definition of Eq.~(\ref{eq:localfNL}).}.
The observational limits on these nonlinearity parameters
are given by~\cite{Komatsu:2010fb}
$ -10 < f_{\rm NL}^{\rm local} < 74
$ for the local type, $ -214 < f_{\rm NL}^{\rm equil} <
266 $ for the equilateral type and $ - 410< f_{\rm
  NL}^{\rm orthog} < 6 $  for the orthogonal type (95\,\% C.L.).

On the other hand,
it has been discussed that there could be a lot of sources of CMB bispectrum other than the primordial non-Gaussianity,
e.g., nonlinear evolution of the temperature fluctuations,
cosmic string and so on.
Primordial magnetic fields (PMFs) may also become the source of
the CMB temperature anisotropies and give a highly non-Gaussian contribution.
In Refs.~\cite{Brown:2005kr,Seshadri:2009sy,Caprini:2009vk,Cai:2010uw, Trivedi:2010gi},
the authors investigated the contribution to the bispectrum of the CMB temperature fluctuations
from the scalar mode PMFs and
roughly estimated the limit of the amplitude of the PMFs. 
As is well known, however,
PMFs excite not only the scalar fluctuation
but also the vector and tensor fluctuations.
In particular, it has been known that the vector contribution may dominate the
scalar one on small scales.
Hence, future CMB experiments, for example, the Planck satellite~\cite{:2006uk},
are expected to give a tighter constraint of the amplitude of the
PMFs from the vector contribution
induced from the magnetic fields.
In this paper, 
we investigate the CMB bispectrum of the vector perturbation
induced from the PMFs
without neglecting the angular dependence.

This paper is organized as follows.
In the next section, we give the formula of the vector bispectrum induced
from PMFs.
Then, in Sec. \ref{sec:result}, we show our result of the CMB bispectrum
from the PMFs and estimate the limit
of the amplitude of the magnetic fields.
We also discuss the shape of the bispectrum in $\ell$ space. 
The final section is devoted to a summary and discussion of this paper.

\section{Formulation of the vector bispectrum induced from PMFs}
\label{sec:formula}

Let us consider the stochastic PMFs $B^b({\bf x}, \tau)$
on the homogeneous background Universe which is characterized by the Friedmann-Robertson-Walker metric
\begin{eqnarray}
ds^2 = a(\tau)^2 \left[ - d\tau^2 + \delta_{bc} dx^b dx^c \right]~,
\end{eqnarray}
where $\tau$ is a conformal time and $a$ is a scale factor.
The expansion of the Universe makes the amplitude of the magnetic fields
decay as $1 / a^2$ and hence we can draw off the time dependence as $B^b({\bf x}, \tau) = 
 B^b({\bf x}) / a^2$.
Then the energy momentum tensor of the PMFs are given by
\begin{eqnarray}
\begin{split}
T^0_{~0} &= - \rho_B = - \frac{1}{8 \pi a^4} B^2({\bf x}) \equiv - \rho_\gamma \Delta_B~, \\
T^0_{~c} &= T^b_{~0} = 0~, \\
T^b_{~c} &= \frac{1}{4\pi a^4} \left[\frac{B^2({\bf x})}{2}\delta^b_{~c} -  B^b({\bf x})B_c({\bf x})\right] \\
&\equiv \rho_\gamma \left(\Delta_B \delta^b_{~c} + \Pi^b_{Bc} \right)~, \label{eq:EMT_PMF} 
\end{split}
\end{eqnarray}
where 
$B^2 = B^b B_b$, and we introduce the photon energy density $\rho_{\gamma}$ in order to include the time dependence; $a^{-4}$.
In the following discussion, 
the index is lowered by $\delta_{bc}$, and the summation is implied for repeated indices.

\subsection{Vector anisotropic stress fluctuations induced from the stochastic magnetic fields}
\label{subsec;vector}

Here, we focus on the vector contribution induced from the PMFs,
which comes from the anisotropic stress of the energy momentum tensor, i.e., $\Pi_{Bab}$.
The Fourier component of $\Pi_{B ab}$ is given by the convolution of the PMFs as
\begin{eqnarray}
\Pi_{B ab}({\bf k}) = -\frac{1}{4\pi \rho_{\gamma,0}}
\int \frac{d^3 {\bf k}'}{(2\pi)^3} B_a({\bf k}') B_b ({\bf k} - {\bf k}')~,
\label{eq:EMT}
\end{eqnarray}
where $\rho_{\gamma, 0}$ denotes the present energy density of photons.
Assuming that $B^a({\bf x})$ is a Gaussian field,
the power spectrum of the PMFs $P_B(k)$ is defined by
\begin{eqnarray}
\langle B_a({\bf k})B_b({\bf p})\rangle
= (2\pi)^3 {P_B(k) \over 2} P_{ab}(\hat{\bf k}) \delta ({\bf k} + {\bf p})~,
\end{eqnarray}  
with a projection tensor
\begin{eqnarray} 
P_{ab}(\hat{\bf k}) 
\equiv \sum_{\sigma = \pm 1} \epsilon^{(\sigma)}_a \epsilon^{(-\sigma)}_b 
 =  \delta_{ab} - \hat{k}_a \hat{k}_b ~,
\end{eqnarray}
which comes from the divergenceless of the PMF.
Here $\hat{\bf k}$ denotes a unit vector, $\epsilon_a^{(\pm 1)}$ is a normalized divergenceless polarization vector which satisfies the orthogonal condition; $\hat{k}^a \epsilon_a^{(\pm 1)} = 0$, and $\sigma (= \pm 1)$ expresses the helicity of the polarization vector.
Although the form of the power spectrum $P_B(k)$ is strongly dependent on the production mechanism,
we, here, assume a simple power law shape given by
\begin{eqnarray}
P_B(k) = {(2\pi)^{n_B+5} \over \Gamma(n_B / 2 + 3/2)k_{1 {\rm Mpc}}^3}
B_{1 {\rm Mpc}}^2 \left({k \over k_{1 {\rm Mpc}}} \right)^{n_B}~,
\end{eqnarray}
where $B_{1 {\rm Mpc}}$ denotes the magnetic field amplitude smoothed on a scale $1 {\rm Mpc}$
and $n_B$ is a spectral index.

The vector anisotropic stress fluctuations are defined with a polarization vector as
\begin{eqnarray}
\Pi_{Bv}^{({\pm 1})}({\bf k}) = \hat{k}_a \epsilon_b^{({\mp 1})} \Pi_{B ab}({\bf k})~.
\label{eq:vecani}
\end{eqnarray}

\subsection{Bispectrum of the vector anisotropic stress fluctuations}
\label{subsec:bispectrum}

As we have mentioned in the previous subsection,
the statistics of the magnetic fields $B^a$ are assumed to be a Gaussian.
Hence 
one can easily find that
the statistics of the vector anisotropic stress fluctuations given by Eq.~(\ref{eq:vecani})
is highly non-Gaussian and the bispectrum
which is corresponding to a Fourier component of 3-point correlation function
has a nonzero value. 

Using the above equations, the bispectrum of $\Pi_{Bv}^{(\pm 1)}({\bf
k})$ is symmetrically formed as
\begin{widetext}
\begin{eqnarray}
\Braket{\prod_{i=1}^{3} \Pi_{Bv}^{(\lambda_i)}({\bf k_i})}
&=& \left[ \prod_{i=1}^3 \int d^3 {\bf k_i'} P_B(k'_i) \right]
\delta({\bf k_1} - {\bf k_1'} + {\bf k_3'}) 
\delta({\bf k_2} - {\bf k_2'} + {\bf k_1'}) 
\delta({\bf k_3} - {\bf k_3'} + {\bf k_2'})
/ \left(- 4 \pi \rho_{\gamma, 0}\right)^{3}
\nonumber \\
&& \times \frac{1}{8} \hat{k_1}_a \epsilon^{(- \lambda_1)}_b 
\hat{k_2}_c \epsilon^{(- \lambda_2)}_d
\hat{k_3}_e \epsilon_f^{(- \lambda_3) }
[P_{ad}(\hat{\bf k_1'}) P_{be}(\hat{\bf k_3'}) P_{cf}(\hat{\bf k_2'}) + \{a \leftrightarrow b \ {\rm or} \ c \leftrightarrow d \ {\rm or} \ e \leftrightarrow f\}]  ~,
\label{eq:3-point}
\end{eqnarray}
\end{widetext}
where $\lambda$ means two helicities: $\lambda_1, \lambda_2, \lambda_3 = \pm 1$ and the curly bracket denotes the symmetric 7 terms under the permutations of indices: $a \leftrightarrow b$, $c \leftrightarrow d$, or $e \leftrightarrow f$.

\subsection{CMB temperature bispectrum}

The CMB temperature fluctuation is expanded into spherical harmonics as
${\Delta T (\hat{\bf n}) \over T} = \sum_{\ell m} a_{\ell m} Y_{\ell m}(\hat{\bf n})$.
The $a_{\ell m}$ sourced by $\Pi_{Bv}^{(\pm 1)}$ can be expressed as 
\begin{eqnarray}
a_{\ell m} = (-i)^{\ell} \int {k^2 dk \over 2\pi^2} \mathcal{T}_{\ell}(k) \sum\limits_{\lambda = \pm 1}
\lambda \Pi_{Bv,\ell m}^{(\lambda)}(k) ~,
\end{eqnarray}
where $\mathcal{T}_{\ell}(k)$ denotes the transfer function of the vector-compensated magnetic mode \cite{Lewis:2004ef, Shaw:2009nf} and we have expanded $\Pi_{Bv}^{(\pm 1)}$ in terms of the spin-$1$ spherical harmonic function ${}_{\mp 1}Y_{\ell m}$ as 
\begin{eqnarray}
\Pi_{Bv,\ell m}^{({\pm 1})}(k) \equiv \int d^2 \hat{\bf k} \Pi_{Bv}^{({\pm 1})}({\bf k}) 
{}_{\mp 1}Y^*_{\ell m}(\hat{\bf k})~.
\end{eqnarray}
In Ref.~\cite{Shiraishi:2010sm}, one can find a more general formulation of $a_{\ell m}$ generated from  vector or tensor perturbations. 
The CMB temperature bispectrum induced from the
vector anisotropic stress $\Pi_{Bv}^{(\pm 1)}$ is given by 
\begin{eqnarray}
\Braket{\prod_{j=1}^3 a_{\ell_j m_j}} 
&=& \left[ \prod\limits^3_{j=1}
(-i)^{\ell_j}
\int {k_j^2 dk_j \over 2\pi^2}
\mathcal{T}_{\ell_j}(k_j)
\sum\limits_{\lambda_j = \pm 1} \lambda_j 
\right]  
\nonumber \\
&&\times
\Braket{ \prod_{j=1}^3
\Pi_{Bv,\ell_j m_j}^{(\lambda_j)}(k_j) 
}.
\label{eq:result}
\end{eqnarray}

In order to calculate the bispectrum of $\Pi^{(\pm 1)}_{B v, \ell
 m}$ derived from Eq. (\ref{eq:3-point}) explicitly, we firstly rewrite all angular dependencies in Eq. (\ref{eq:3-point}) in terms of 
the spin-weighted spherical harmonics as
\begin{eqnarray}
&& \hat{r_1}_a \epsilon^{(\mp 1)}_b(\hat{\bf r_2}) P_{ab} (\hat{\bf r_3}) 
= \mp \sum_{\sigma = \pm 1} \sum_{m  m'} \left(\frac{4 \pi}{3}\right)^2 
Y_{1 m} (\hat{\bf r_1}) \nonumber \\
&&\qquad \times {}_{\mp 1}Y_{1 m'} (\hat{\bf r_2}) 
{}_{- \sigma}Y^*_{1 m} (\hat{\bf r_3}) {}_{\sigma}Y^*_{1 m'} (\hat{\bf r_3})~, \\
&&  \delta\left( \sum_{i=1}^3 {\bf r_i} \right) 
= 8 \int_0^\infty y^2 dy 
\left[ \prod_{i=1}^3 \sum_{l_i m_i} 
 (-1)^{l_i/2} j_{l_i}(r_i y)
  \right. \nonumber \\ 
&& \left. \qquad  \times  
Y_{l_i m_i}(\hat{\bf r_i}) \right]
I_{l_1 l_2 l_3}^{0~0~0}
\left(
  \begin{array}{ccc}
  l_1 & l_2 & l_3 \\
  m_1 & m_2 & m_3 
  \end{array}
 \right)
 \label{delta3}  ~, 
\end{eqnarray}
where the matrix means the Wigner-$3j$ symbol, $j_l(x)$ is the spherical Bessel function and
\begin{eqnarray}
I^{s_1 s_2 s_3}_{l_1 l_2 l_3} 
\equiv \sqrt{\frac{(2 l_1 + 1)(2 l_2 + 1)(2 l_3 + 1)}{4 \pi}}
\left(
  \begin{array}{ccc}
  l_1 & l_2 & l_3 \\
   s_1 & s_2 & s_3 
  \end{array}
 \right)~. \nonumber 
\end{eqnarray} 
Then, using the fact that the angular integrals of the spin spherical harmonics in Eq.~(\ref{eq:result}) can be expressed with respect to the Wigner-$3j$ symbols as 
\begin{eqnarray}
&& \int d^2 \hat{\bf r} 
\prod_{i=1}^4 {}_{s_i} Y_{l_i m_i} (\hat{\bf r}) 
= \sum_{l_5 m_5 s_5} I_{l_1~ l_2~ l_5}^{-s_1 -s_2 ~ s_5} I_{l_3~ l_4~ l_5}^{-s_3 -s_4 -s_5} 
\nonumber \\
&&\qquad \times  
\left(
  \begin{array}{ccc}
  l_1 & l_2 & l_5 \\
  m_1 & m_2 & m_5 
  \end{array}
 \right)  
\left(
  \begin{array}{ccc}
  l_3 &  l_4 & l_5 \\
  m_3 & m_4 & m_5 
  \end{array}
 \right)~,
\label{eq:ang_int_kdpdqd}
\end{eqnarray}
and summing up these Wigner-$3j$ symbols over the azimuthal quantum
numbers as \cite{Hu:2001fa,Jahn/Hope:1954}
\begin{eqnarray}
&& \sum_{\substack{m_4 m_5 m_6 \\ m_7 m_8 m_9}} 
\left(
  \begin{array}{ccc}
  l_4 & l_5 & l_6 \\
  m_4 & m_5 & m_6 
  \end{array}
 \right)
\left(
  \begin{array}{ccc}
  l_7 & l_8 & l_9 \\
  m_7 & m_8 & m_9 
  \end{array}
 \right) \nonumber \\
&& \times
\left(
  \begin{array}{ccc}
  l_4 & l_7 & l_1 \\
  m_4 & m_7 & m_1 
  \end{array}
 \right)
\left(
  \begin{array}{ccc}
  l_5 & l_8 & l_2 \\
  m_5 & m_8 & m_2
  \end{array}
 \right)
\left(
  \begin{array}{ccc}
  l_6 & l_9 & l_3 \\
  m_6 & m_9 & m_3 
  \end{array}
 \right) \nonumber \\ 
&& \qquad\qquad = \left(
  \begin{array}{ccc}
  l_1 & l_2 & l_3 \\
  m_1 & m_2 & m_3
  \end{array}
 \right)
\left\{
  \begin{array}{ccc}
  l_1 & l_2 & l_3 \\
  l_4 & l_5 & l_6 \\
  l_7 & l_8 & l_9 
  \end{array}
 \right\}~, \\
&& \sum_{m_4 m_5 m_6} (-1)^{\sum_{i=4}^6 l_i - m_i}
\left(
  \begin{array}{ccc}
  l_5 & l_1 & l_6 \\
  m_5 & -m_1 & -m_6 
  \end{array}
 \right) \nonumber \\
&&\qquad \times
\left(
  \begin{array}{ccc}
  l_6 & l_2 & l_4 \\
  m_6 & -m_2 & -m_4 
  \end{array}
 \right)
\left(
  \begin{array}{ccc}
  l_4 & l_3 & l_5 \\
  m_4 & -m_3 & -m_5 
  \end{array}
 \right) \nonumber \\
 &&\qquad\qquad = \left(
  \begin{array}{ccc}
  l_1 & l_2 & l_3 \\
  m_1 & m_2 & m_3 
  \end{array}
 \right) 
\left\{
  \begin{array}{ccc}
  l_1 & l_2 & l_3 \\
  l_4 & l_5 & l_6 
  \end{array}
 \right\}~,
\end{eqnarray}
we obtain the final form as
\begin{widetext}
\begin{eqnarray}
\Braket{\prod_{i=1}^3 \Pi^{(\lambda_i)}_{B v, \ell_i m_i}(k_i) } 
&=& \left(%
\begin{array}{ccc}
  \ell_1  & \ell_2    & \ell_3   \\
   m_1    & m_2       & m_3   \\
\end{array}%
\right)
\left[ \prod_{i=1}^3 \int_0^{k_D} k_i'^2 dk_i' P_B(k_i') \right]
/(-4 \pi \rho_{\gamma,0})^3
\nonumber \\ 
&& \times 
\sum_{L L' L''} \sum_{S, S', S'' = \pm 1} 
\left\{
  \begin{array}{ccc}
  \ell_1 & \ell_2 & \ell_3 \\
  L' & L'' & L 
  \end{array}
 \right\}
f^{S'' S \lambda_1}_{L'' L \ell_1}(k_3',k_1',k_1) f^{S S' \lambda_2}_{L L' \ell_2}(k_1',k_2',k_2)
f^{S' S'' \lambda_3}_{L' L'' \ell_3}(k_2',k_3',k_3), 
 \label{eq:calF_exact} \\ 
f^{S'' S \lambda}_{L'' L \ell}(r_3, r_2, r_1) 
&\equiv& \frac{2(8\pi)^{3/2}}{3}
\sum_{L_1 L_2 L_3} \int_0^\infty y^2 dy j_{L_3}(r_3 y) j_{L_2}(r_2 y) j_{L_1}(r_1 y)  \nonumber \\
&& \times
 \lambda (-1)^{\ell + L_2+L_3} (-1)^{\frac{L_1 + L_2 + L_3}{2}} 
I^{0~0~0}_{L_1 L_2 L_3} I^{0 S'' -S''}_{L_3 1 L''} I^{0 S -S}_{L_2 1 L} 
I_{L_1 \ell 2}^{0 \lambda -\lambda} 
 \left\{
  \begin{array}{ccc}
  L'' & L & \ell \\
  L_3 & L_2 & L_1 \\
  1 & 1 & 2
  \end{array}
 \right\}~, \label{eq:f_exact}
\end{eqnarray}
\end{widetext}
where $k_D$ is the is the Alfv\'en-wave damping length scale and the $2
\times 3$ and $3 \times 3$ matrices of a curly bracket denote the
Wigner-$6j$ and $9j$ symbols, respectively. 
From the above full expression, we can easily find 
that the dependence on $m_1 ,m_2$ and $m_3$
is confined only to the first Wigner-$3j$ symbol. This guarantees the
rotational invariance of the bispectrum, hence, we can define the
angle-averaged quantity $B_{\ell_1 \ell_2 \ell_3}$ like the nonmagnetic scalar case \cite{Komatsu:2001rj} as
\begin{eqnarray}
\Braket{\prod_{i=1}^3 a_{\ell_i m_i}} 
 &\equiv& \left(%
\begin{array}{ccc}
  \ell_1  & \ell_2    & \ell_3   \\
   m_1    & m_2       & m_3   \\
\end{array}%
\right) B_{\ell_1 \ell_2 \ell_3}~.
\end{eqnarray}
Detailed derivation and more expansion of this equation will be shown in a full paper~\cite{Shiraishi:inprep_PMF}.
\section{Results}
\label{sec:result}

Here we show the result of the CMB temperature bispectrum
induced from the vector anisotropic stress $\Pi_{Bv}^{(\lambda)}$.

\begin{figure}[t]
  \begin{center}
    \includegraphics{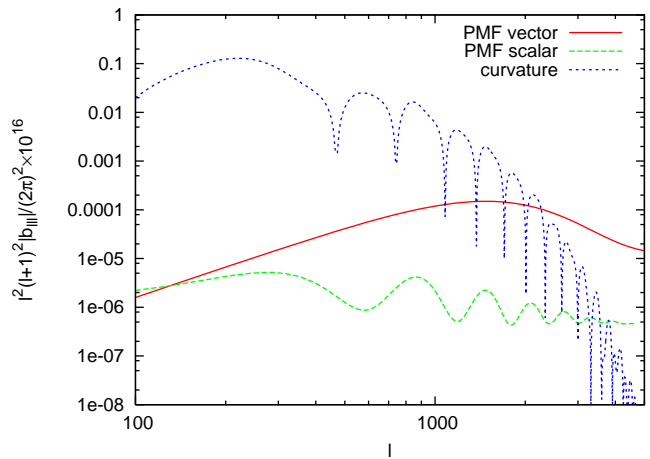}
  \end{center}
  \caption{(color online). Absolute values of the normalized reduced bispectra of temperature fluctuation 
for a configuration $\ell_1 = \ell_2 = \ell_3$. The lines correspond to
 the spectra generated from vector anisotropic stress (red solid line),
 scalar isotropic stress (green dashed line) and primordial
 non-Gaussianity defined in Eq.~(\ref{eq:localfNL}) with $f^{\rm local}_{\rm NL} =
 5$ (blue dotted line). The PMF parameters are fixed to ($n_B = -2.9$ and $B_{1 {\rm Mpc}} = 4.7 {\rm nG})$ and the other cosmological parameters are fixed to the mean values limited from the WMAP-7 yr data reported in Ref. \cite{Komatsu:2010fb}.
  }
  \label{fig:vec_III_samel.eps}
\end{figure}

In Fig.~\ref{fig:vec_III_samel.eps},
we show the reduced bispectra of temperature fluctuation
induced by the PMFs 
defined as~\cite{Komatsu:2001rj}
$b_{\ell_1 \ell_2 \ell_3}
\equiv 
 (I^{0~0~0}_{\ell_1 \ell_2 \ell_3})^{-1} B_{\ell_1 \ell_2 \ell_3}$,
for $\ell_1 = \ell_2 = \ell_3$. 
The red solid line is corresponding to the bispectrum given by
Eq.~(\ref{eq:result}). One can see that the overall behavior, 
such as the peak location at $\ell \sim 1500$, is very
similar to that of the angular power spectrum $C_\ell$ from the vector mode
as calculated in Ref. \cite{Shaw:2009nf}. 
The amplitude is smaller than $C_\ell^{3/2}$ 
which is expected by a naive order of magnitude estimate. It is because the configuration of multipoles, corresponding to the angular dependence of wave
number vectors, is limited to the conditions placed by the Wigner symbols. 
By this effect, $b_{\ell \ell \ell}$ is suppressed by a factor
$\ell^{-2}$ from $C_\ell^{3/2}$. 
This is the reason that our constraint from the vector bispectrum on
the PMF is not much stronger than expected from the scalar counterpart.

For comparison, we also compute the CMB bispectrum sourced from scalar mode
isotropic stress of the PMFs [$\Delta_B$ of Eq. (\ref{eq:EMT_PMF})]. The amplitude at large scale is consistent
with the result of previous study \cite{Seshadri:2009sy}
although we newly consider the full angular dependence. From this
figure, we can confirm that in the bispectrum, the vector mode dominates at small scales, specifically $\ell \sim 1500$, in the same manner as the angular power spectrum.

\begin{figure}[t]
  \begin{center}
    \includegraphics{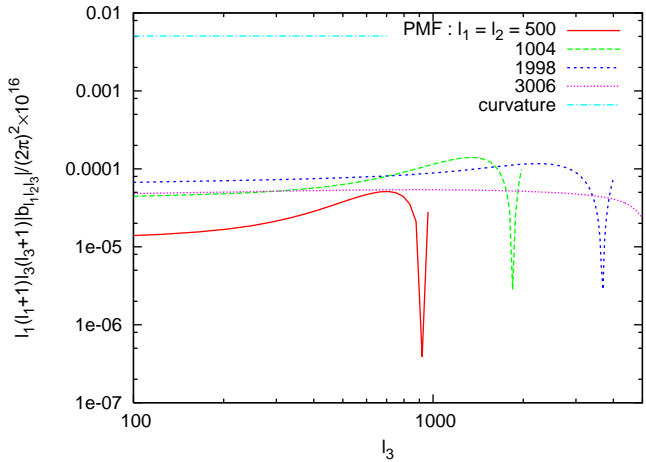}
  \end{center}
  \caption{(color online). Absolute values of the normalized reduced
 bispectra of temperature fluctuation given by Eq.~(\ref{eq:result}) and
 generated from primordial non-Gaussianity given by
 Eq. (\ref{eq:CMB_bis_local}) as a function of $\ell_3$ with $\ell_1$ and $\ell_2$ fixed to some values as indicated.
Each parameter is fixed to the same values defined in Fig. \ref{fig:vec_III_samel.eps}.
}
  \label{fig:vec_III_difl.eps}
\end{figure}

In Fig.~\ref{fig:vec_III_difl.eps}, we also show the reduced bispectrum  $b_{\ell_1 \ell_2 \ell_3}$ with respect to $\ell_3$ with setting $\ell_1 = \ell_2$.
From this figure,
we can see that the normalized reduced bispectrum of the vector magnetic mode is nearly flat as 
\begin{eqnarray}
\ell_1 (\ell_1+1) \ell_3 (\ell_3+1) |b_{\ell_1 \ell_2 \ell_3}| 
\sim 2 \times 10^{-19} 
\left(\frac{B_{1 \rm Mpc}}{4.7 \rm nG} \right)^6 ,
\end{eqnarray}
and it is clear that $b_{\ell_1 \ell_2 \ell_3}$ for $n_B = -2.9$ dominates in $\ell_1 = \ell_2 \gg \ell_3$.
This seems to mean the shape of CMB bispectrum generated from vector anisotropic stress of PMF is close to the local-type configuration if the power spectrum of PMF is nearly scale invariant. 
Therefore, in order to obtain a valid constraint on the magnitude of PMF, 
the bispectrum induced from PMF should be compared with that from the
local-type primordial non-Gaussianity defined in
Eq.~(\ref{eq:local_pri_bis}), which is estimated as \cite{Riot:2008ng}
\begin{eqnarray}
\ell_1 (\ell_1+1) \ell_3 (\ell_3+1)b_{\ell_1 \ell_2 \ell_3} \sim 4
 \times 10^{-18} f^{\rm local}_{\rm NL}~. \label{eq:CMB_bis_local}
\end{eqnarray}
From these equations, the relation between the magnitudes of the PMF and $f^{\rm local}_{\rm NL}$ is derived as
\begin{eqnarray}
\left( \frac{B_{1\rm Mpc}}{1\rm nG} \right) \sim 7.74~ 
|f^{{\rm local}}_{\rm NL}|^{1/6} \ \ ({\rm for ~ n_B = -2.9})~.
\end{eqnarray} 

By making use of the above relation, we can place the upper bound of strength of the PMF.
If we assume $|f^{\rm local}_{\rm NL}| < 100$ as considered in Ref.~\cite{Seshadri:2009sy}, we obtain $B_{1\rm Mpc} < 17{\rm nG}$,
which is stronger by a factor of 2 than estimated from the scalar bispectrum in Ref. \cite{Seshadri:2009sy}. On the other hand, from current observational lower bound from the WMAP 7-yr data mentioned in Sec. \ref{sec:intro}, namely $f_{\rm NL}^{\rm local} > -10$, we derive $B_{1\rm Mpc} < 11{\rm nG}$. Furthermore, if we use the observational data of the PLANCK experiment \cite{:2006uk} which is expected to reach $|f_{\rm NL}^{\rm local}| < 5$, we will meet $B_{1\rm Mpc} < 10 {\rm nG}$. 

\section{Summary and Discussion}
\label{sec:sum}

In this paper we present a bispectrum of CMB
temperature fluctuation induced from the vector mode of PMFs by taking
into account the full angular dependence of the bispectrum of magnetic
fields. It is found that the CMB bispectrum from the magnetic vector
mode dominates at small scales compared to that from the magnetic scalar
mode which has been calculated in the literature. It is also found that
the bispectrum has significant signals on the squeezed limit, namely, the
local-type configuration, if the magnetic field power spectrum is nearly
scale invariant.

So far, constraints on the PMFs have been derived mainly from the power
spectrum of CMB, dark matter clustering, and the structure
formation (e.g. \cite{Paoletti:2010rx, 2010arXiv1006.4242S}). 
However, it has been
reported that the PMF parameters have some degeneracy with the other
cosmological parameters, such as neutrino masses
\cite{Yamazaki:2010jw}. From this fact and the non-Gaussian nature
of the magnetic field induced fluctuations, the CMB
bispectrum will give us supplementary and complementary information
about the PMF parameters.  By translating the current bound on the
local-type non-Gaussianity from the CMB bispectrum into the bound on the
amplitude of the magnetic fields, we obtain a new limit: $B_{1\rm Mpc} <
11 {\rm nG}$. This is a rough estimate and a tighter constraint is
expected if one considers the full $\ell$ contribution by using an
appropriate estimator of the magnetic field bispectrum.

Because of the complicated discussions and mathematical manipulations, here
we only show the temperature bispectrum from the vector mode of the
PMFs. A full treatment of the bispectra of
CMB temperature and polarization from the vector and scalar modes will be
presented elsewhere.

\begin{acknowledgments}
This work is supported by Grant-in-Aid for JSPS Research under
 Grant No.~22-7477 (M. S.), and JSPS Grant-in-Aid for Scientific
Research under Grant Nos.~22340056 (S. Y.), 21740177, 22012004 (K. I.),
 and 21840028 (K. T.).
This work is supported in part by the Grant-in-Aid for Scientific
Research on Priority Areas No. 467 "Probing the Dark Energy through an
 Extremely Wide and Deep Survey with Subaru Telescope" and by the
 Grant-in-Aid for Nagoya University Global COE Program, "Quest for
 Fundamental Principles in the Universe: from Particles to the Solar
 System and the Cosmos," from the Ministry of Education, Culture,
 Sports, Science and Technology of Japan. 
\end{acknowledgments}

\bibliography{paper}
\end{document}